\documentstyle[aps, preprint]{revtex}

\begin{document}
\draft
\title{Convergence of energy-dependent incommensurate
antiferromagnetic neutron scattering peaks to commensurate
resonance in underdoped bilayer cuprates}
\author{Shiping Feng and Feng Yuan}
\address{Department of Physics, Beijing Normal University, Beijing
100875, China\\
National Laboratory of Superconductivity, Academia Sinica, Beijing
100080, China}
\author{Zhao-Bin Su}
\address{Institute of Theoretical Physics and Interdisciplinary
Center of Theoretical Studies, Chinese Academy of Sciences,
Beijing 100080, China}
\author{Lu Yu}
\address{The Abdus Salam International Centre for Theoretical
Physics, 34014 Trieste, Itlay\\
Institute of Theoretical Physics and Interdisciplinary Center of
Theoretical Studies, Chinese Academy of Sciences, Beijing 100080,
China}
\maketitle
\date{\today}

\begin{abstract}
The recently discovered coexistence of incommensurate
antiferromagnetic neutron scattering peaks and  commensurate
resonance in underdoped YBa$_2$Cu$_3$O$_{6+x}$ is calling for an
explanation. Within the $t$-$J$ model, the doping and energy
dependence of the spin dynamics of the underdoped bilayer cuprates
in the normal state is studied based on the fermion-spin theory by
considering the bilayer interactions. Incommensurate peaks are
found at $[(1\pm\delta)\pi,\pi] $ and $[\pi,(1\pm\delta)\pi]$ at
low energies with $\delta$ initially increasing with doping at
low dopings and then saturating at higher dopings. These
incommensurate peaks are suppressed, and the parameter $\delta$ is
reduced with increasing energy. Eventually  it converges to the
$[\pi,\pi]$ resonance peak. Thus  the recently observed
coexistence is interpreted in terms of bilayer interactions.
\end{abstract}




The interplay between antiferromagnetism (AF) and
superconductivity (SC) in high $T_c$ cuprates is well-established
by now,\cite{rev} but its full understanding is still a
challenging issue. Experimentally the inelastic neutron scattering
(INS) can provide rather detailed information on the spin dynamics
of doped single layer and bilayer cuprates.
\cite{yamada,aeppli,ros91,res,resunder,bsco,ybic,dai01,bour} An
important issue is whether the behavior of AF fluctuations in
these compounds is universal or not.  A distinct feature of
single layer La$_{2-x}$Sr$_x$CuO$_4$ (LSCO) is the presence of
incommensurate antiferromagnetic (ICAF) peaks at low energy
INS, {\it i.e.,} the AF scattering peaks are shifted from the AF
wave vector [$\pi $,$\pi $] to four points [$\pi (1\pm\delta ),
\pi $] and [$\pi ,(1\pm \delta )\pi $] (in units of inverse
lattice constant) with $\delta $ as the incommensurability (IC)
parameter, which depends on doping concentration, but not on
energy. Moreover, ICAF is observed both above and below $T_c$ in
the entire range of doping, from underdoped to overdoped samples.
\cite{rev,yamada,aeppli} In contrast, a sharp resonance peak
(around 41 meV) is observed in optimally doped bilayer
YBa$_2$Cu$_3$O$_{6+x}$ (YBCO) at the commensurate AF wave vector
[$\pi $,$\pi $] in the SC state.\cite{ros91,res} Such a resonance
has also been observed in underdoped YBCO samples with resonance
energy scaling down with the SC $T_c$, being present both below
and above $T_c$.\cite{resunder} Recently, this resonance peak has
been observed in another class of bilayer SC cuprates
Bi$_2$Sr$_2$CaCu$_2$O$_{8+\delta}$ (BSCO). \cite{bsco} Such a peak
has, however, never been observed in LSCO. A very important  new
development is the observation of ICAF in underdoped YBCO in both
SC and normal states, with INS pattern and doping dependence being
very similar (linear in doping for low dopings) to that of LSCO.
\cite{ybic,dai01} However, the IC peak position is energy
dependent in underdoped YBCO.  A challenging issue for theory is
to explain the coexistence of this energy-dependent ICAF
scattering and commensurate resonance in bilayer cuprates.

Theoretically the ICAF has been interpreted, among others, in
terms of Fermi surface nesting \cite{nesting,earthe} or stripe
formation. \cite{stripe} The energy dependence of IC parameter
$\delta $ on energy for underdoped YBCO makes the stripe
interpretation rather difficult to accept. On the other hand, the
commensurate resonance peak has been interpreted as due to spin-1
collective (particle-hole) excitations,\cite{earthe,latthe,brink}
or particle-particle excitations, \cite{so5} or interlayer
tunneling. \cite{yin} These theoretical treatments are mostly
addressing the SC state, and heavily rely on adjusting band
structure parameters, like the next nearest neighbor hopping
$t^{\prime}$, {\it etc}. To the best of our knowledge, the ICAF
and commensurate resonance peak in underdoped bilayer cuprates
have not yet been treated from a unified point of view for the
normal state. No explicit predictions on doping and energy
dependence of the ICAF peaks have been made so far.

In our earlier work using the fermion-spin theory, \cite{sp94}
the dynamical spin structure factor (DSSF) has been calculated
for LSCO within the single layer $t$-$J$ model \cite{yuan}, and
the obtained doping dependence of the IC parameter $\delta $ is
consistent with experiments. \cite{rev,yamada,aeppli} In this
paper we show explicitly if the bilayer interactions are included,
one can reproduce all main features in the normal state observed
experimentally on YBCO,\cite{ybic,dai01} including the doping
dependence of ICAF at low energies and $[\pi,\pi ]$ resonance at
relatively high energy. The bilayer band splitting in BSCO has
been observed in the angle-resolved photoemission spectroscopy in
both normal and superconducting states. \cite{feng} The
convergence of ICAF peaks at lower energies to commensurate
resonance peak at higher energy is rather similar to the scenario
argued in Ref. \cite{bour} for the SC state, and the DSSF we
derive from the simple  $t$-$J$ model (without additional terms
and adjustable parameters) demonstrates explicitly this
convergence.

The  $t$-$J$ model in bilayer structures is expressed as, 
\begin{eqnarray}
H&=&-t\sum_{ai\hat{\eta}\sigma}C_{ai\sigma}^{\dagger}
C_{ai+\hat{\eta}\sigma}-t_{\perp}\sum_{i\sigma}
(C_{1i\sigma}^{\dagger}C_{2i\sigma}+{\rm h.c.})-\mu\sum_{ai\sigma}
C_{ai\sigma }^{\dagger }C_{ai\sigma } \nonumber \\
&+&J\sum_{ai\hat{\eta}}{\bf S}_{ai}\cdot {\bf S}_{ai+\hat{\eta}}+
J_{\perp}\sum_i{\bf S}_{1i}\cdot {\bf S}_{2i},~~~
\end{eqnarray}
where $\hat{\eta}=\pm\hat{x}$, $\pm\hat{y}$, $a=1,2$ is plane
indices, and ${\bf S}_{ai}=C_{ai}^{\dagger}{\vec{\sigma}}C_{ai}/2$
are spin operators with ${\vec{\sigma}}=(\sigma_x,\sigma_y,
\sigma_z)$ as Pauli matrices. The $t$-$J$ Hamiltonian is
supplemented by the single occupancy local constraint
$\sum_\sigma C_{ai\sigma }^{\dagger }C_{ai\sigma }\leq 1$. This
local constraint can be treated {\it properly in analytical form}
within the fermion-spin theory \cite{sp94} based on the slave
particle approach,
\begin{eqnarray}
C_{ai\uparrow}=h_{ai}^{\dagger}S_{ai}^{-}, ~~~~ C_{ai\downarrow}=
h_{ai}^{\dagger}S_{ai}^{+},
\end{eqnarray}
where the spinless fermion operator $h_{ai}$ keeps track of the
charge (holon), while the pseudospin operator $S_{ai}$ keeps track
of the spin (spinon), and the {\it low-energy} Hamiltonian of the
bilayer $t$-$J$ model (1) can be rewritten in the fermion-spin
representation as,
\begin{eqnarray}
H&=&t\sum_{ai\hat{\eta}}h^{\dagger}_{ai+\hat{\eta}}h_{ai}
(S^{+}_{ai}S^{-}_{ai+\hat{\eta}}+S^{-}_{ai}S^{+}_{ai+\hat{\eta}})
+t_{\perp}\sum_{i}(h^{\dagger}_{1i}h_{2i}+h^{\dagger}_{2i}h_{1i})
(S^{+}_{1i}S^{-}_{2i}+S^{-}_{1i}S^{+}_{2i}) \nonumber \\
&+&\mu\sum_{ai}h^{\dagger}_{ai}h_{ai}+J_{{\rm eff}}
\sum_{ai\hat{\eta}}{\bf S}_{ai}\cdot {\bf S}_{ai+\hat{\eta}}
+J_{\perp{\rm eff}}\sum_{i}{\bf S}_{1i}\cdot {\bf S}_{2i},
\end{eqnarray}
with $J_{{\rm eff}}=J[(1-p)^{2}-\phi^{2}]$ and $J_{\perp{\rm eff}}
=J[(1-p)^{2}-\phi^{2}_{\perp}]$, where $p$ is the hole doping
concentration, the holon in-plane and bilayer hopping parameters
$\phi=\langle h^{\dagger}_{ai}h_{ai+\hat{\eta}}\rangle$ and
$\phi_{\perp}=\langle h^{\dagger}_{1i}h_{2i}\rangle$, and
$S^{+}_{ai}$ ($S^{-}_{ai}$) as the pseudospin raising (lowering)
operators. In the bilayer system, because of the two coupled
CuO$_2$ planes, the energy spectrum has two branches. In this
case, the one-particle spinon and holon Green's functions are
matrices, and are expressed as,
\begin{eqnarray}
D(i-j,\tau-\tau^{\prime})&=&D_{L}(i-j,\tau-\tau^{\prime})+\tau_{x}
D_{T}(i-j,\tau-\tau^{\prime}), \nonumber \\
g(i-j,\tau-\tau^{\prime})&=&g_{L}(i-j,\tau-\tau^{\prime})+\tau_{x}
g_{T}(i-j,\tau-\tau^{\prime}),
\end{eqnarray}
respectively, where the longitudinal and transverse parts are
defined as,
\begin{eqnarray}
D_{L}(i-j,\tau-\tau^{\prime})&=&-\langle T_{\tau}S_{ai}^{+}(\tau)
S_{aj}^{-}(\tau^{\prime})\rangle, \nonumber \\
g_{L}(i-j,\tau-\tau^{\prime})&=&-\langle T_{\tau}h_{ai}(\tau)
h_{aj}^{\dagger}(\tau^{\prime})\rangle, \nonumber\\
D_{T}(i-j,\tau-\tau^{\prime})&=&-\langle T_{\tau}S_{ai}^{+}(\tau)
S_{a^{\prime}j}^{-}(\tau^{\prime})\rangle, \nonumber\\
g_{T}(i-j,\tau-\tau^{\prime})&=&-\langle T_{\tau}h_{ai}(\tau)
h_{a^{\prime}j}^{\dagger}(\tau^{\prime})\rangle,
\end{eqnarray}
with $a\neq a^{\prime}$, while $\tau_x$ is the Pauli matrix in
the pseudospin space of the layer index. Within this framework,
the spin fluctuations only couple to spinons, but the strong
correlation between holons and spinons is included
self-consistently through the holon's parameters entering the
spinon propagator. Therefore both spinons and holons are involved
in the spin dynamics. The universal behavior of the integrated
spin response and ICAF in underdoped single layer cuprates have
been discussed within the fermion-spin theory by considering
spinon fluctuations around the mean-field (MF) solution,
\cite{yuan} where the spinon part is treated by the loop expansion
to the second order. Following the previous discussions for the
single layer case, DSSF of bilayer cuprates is obtained explicitly
as,
\begin{eqnarray}
S({\bf k},\omega)&=&-2[1+n_{B}(\omega)][2{\rm Im}D_{L}({\bf k},
\omega)+2{\rm Im}D_{T}({\bf k},\omega)]  \nonumber \\
&=&-{\frac{4[1+n_{B}(\omega)](B_{k}^{(1)})^{2}
{\rm Im}\Sigma_{LT}^{(s)}({\bf k},\omega)}{[\omega^{2}-
(\omega_{k}^{(1)})^{2}-B_{k}^{(1)}{\rm Re}\Sigma_{LT}^{(s)}
({\bf k},\omega)]^{2}+[B_{k}^{(1)}{\rm Im}\Sigma_{LT}^{(s)}
({\bf k},\omega)]^{2}}},
\end{eqnarray}
where the full spinon Green's function,
\begin{eqnarray}
D^{-1}({\bf k},\omega )=D^{(0)-1}({\bf k},\omega)-\Sigma^{(s)}
({\bf k},\omega),
\end{eqnarray}
with the longitudinal and transverse MF spinon Green's functions,
\begin{eqnarray}
D_{L}^{(0)}({\bf k},\omega)&=& 1/2\sum_{\nu}B_{k}^{(\nu)}
/[\omega^{2}-(\omega_{k}^{(\nu)})^{2}], \nonumber \\
D_{T}^{(0)}({\bf k},\omega)&=& 1/2\sum_{\nu}(-1)^{\nu+1}
B_{k}^{(\nu)}/[\omega^{2}-(\omega_{k}^{(\nu)})^{2}],
\end{eqnarray}
respectively, where $\nu=1,2$, and
\begin{eqnarray}
{\rm Im}\Sigma_{LT}^{(s)}({\bf k},\omega)&=&{\rm Im}
\Sigma_{L}^{(s)}({\bf k},\omega)+{\rm Im}\Sigma_{T}^{(s)}
({\bf k},\omega), \nonumber\\
{\rm Re}\Sigma_{LT}^{(s)}({\bf k},\omega)&=&{\rm Re}
\Sigma_{L}^{(s)}({\bf k},\omega)+{\rm Re}\Sigma_{T}^{(s)}
({\bf k},\omega),
\end{eqnarray}
while ${\rm Im}\Sigma_{L}^{(s)}({\bf k},\omega)$ (${\rm Im}
\Sigma_{T}^{(s)}({\bf k},\omega)$) and ${\rm Re}\Sigma_{L}^{(s)}
({\bf k},\omega)$ (${\rm Re}\Sigma_{T}^{(s)}({\bf k},\omega)$)
are the imaginary and real parts of the second order longitudinal
(transverse) spinon self-energy, respectively, obtained from the
holon bubble as,
\begin{eqnarray}
\Sigma_{L}^{(s)}({\bf k},\omega)&=&(1/N)^2\sum_{pp^{\prime}}
\sum_{\nu\nu^{\prime}\nu^{\prime\prime}}\Pi_{\nu\nu^{\prime}
\nu^{\prime\prime}}({\bf k},{\bf p},{\bf p^{\prime}},\omega),
\nonumber\\
\Sigma_{T}^{(s)}({\bf k},\omega)&=&(1/N)^2\sum_{pp^{\prime}}
\sum_{\nu\nu^{\prime}\nu^{\prime\prime}}(-1)^{\nu+\nu^{\prime}+
\nu^{\prime\prime}+1}\Pi_{\nu\nu^{\prime}\nu^{\prime\prime}}
({\bf k},{\bf p},{\bf p^{\prime}},\omega),
\end{eqnarray}
with
\begin{eqnarray}
\Pi_{\nu\nu^{\prime}\nu^{\prime\prime}}({\bf k},{\bf p},
{\bf p^{\prime}},\omega) &=&\left(Zt[\gamma_{p^{\prime}+p+k}+
\gamma_{k-p^{\prime}}]+t_{\perp}[(-1)^{\nu^{\prime}+\nu^{\prime
\prime}}+(-1)^{\nu+\nu^{\prime\prime}}]\right)^{2}
{\frac{B_{k+p}^{(\nu^{\prime\prime})}}
{16\omega_{k+p}^{(\nu^{\prime\prime})}}} \nonumber \\
&\times &\left({\frac{F_{\nu\nu^{\prime}\nu^{\prime\prime}}^{(1)}
({\bf k},{\bf p},{\bf p^{\prime}})}
{\omega+\xi_{p+p^{\prime}}^{(\nu^{\prime})}-
\xi_{p^{\prime}}^{(\nu)}-\omega_{k+p}^{(\nu^{\prime\prime})}}}-
{\frac{F_{\nu\nu^{\prime}\nu^{\prime\prime}}^{(2)}({\bf k},{\bf p},
{\bf p^{\prime }})}{\omega+\xi_{p+p^{\prime}}^{(\nu^{\prime})}-
\xi_{p^{\prime}}^{\nu}+\omega_{k+p}^{(\nu^{\prime\prime})}}}
\right) ,
\end{eqnarray}
where $\gamma_{{\bf k}}=(1/Z)\sum_{\hat{\eta}}e^{i{\bf k}\cdot
\hat{\eta}}$, $Z$ is the coordination number, 
\begin{eqnarray}
&& B_k^{(\nu)}=B_{k} -J_{\perp {\rm eff}}[\chi_{\perp}+
2\chi_{\perp}^z(-1)^{\nu}][\epsilon_{\perp}+(-1)^{\nu}]
\nonumber\\
&& B_{k}=\lambda [(2\epsilon\chi^{z}+\chi)\gamma_{k}-
(\epsilon\chi+2\chi^{z})],\;\;\; \lambda=2ZJ_{{\rm eff}},
\nonumber\\
&& \epsilon=1+2t\phi/J_{{\rm eff}},\;\;\; \epsilon_{\perp}=1+
4t_{\perp}\phi_{\perp}/J_{\perp {\rm eff}}, \nonumber\\
&& F_{\nu\nu^{\prime}\nu^{\prime\prime}}^{(1)}({\bf k},{\bf p},
{\bf p^{\prime }})=n_{F}(\xi_{p+p^{\prime}}^{(\nu^{\prime })})
[1-n_{F}(\xi_{p^{\prime}}^{(\nu)})]-n_{B}
(\omega_{k+p}^{(\nu^{\prime\prime})})
[n_{F}(\xi_{p^{\prime}}^{(\nu)})-
n_{F}(\xi_{p+p^{\prime}}^{(\nu^{\prime})})], \nonumber\\
&& F_{\nu\nu^{\prime}\nu^{\prime\prime}}^{(2)}({\bf k},{\bf p},
{\bf p^{\prime}})=n_{F}(\xi_{p+p^{\prime}}^{(\nu^{\prime})})
[1-n_{F}(\xi_{p^{\prime}}^{(\nu)})]+[1+n_{B}
(\omega_{k+p}^{(\nu^{\prime\prime})})]
[n_{F}(\xi_{p^{\prime }}^{(\nu)})-
n_{F}(\xi_{p+p^{\prime}}^{(\nu^{\prime})})],
\end{eqnarray}
$n_{F}(\xi_{k}^{(\nu)})$ and $n_{B}(\omega_{k}^{(\nu)})$ are the
fermion and boson distribution functions, respectively, and the
MF holon and spinon excitations,
\begin{eqnarray}
\xi_{k}^{(\nu)}&=&2Zt\chi\gamma_{k}+\mu+2\chi_{\perp}
t_{\perp}(-1)^{\nu+1}, \nonumber\\
(\omega_{k}^{(\nu)})^2&=&\omega_{k}^2+\Delta_{k}^2(-1)^{\nu+1},
\end{eqnarray}
with $\omega_{k}^2=A_{1}\gamma_{k}^{2}+A_{2}\gamma_{k}+A_{3}$,
$\Delta_{k}^{2}=X_{1}\gamma_{k}+X_{2}$, 
\begin{eqnarray}
A_{1}=&&\alpha\epsilon\lambda^{2}(\chi/2+\epsilon\chi^{z}),
\nonumber\\
A_{2}=&&\epsilon\lambda^{2}[(1-Z)\alpha(\epsilon\chi/2+\chi^{z})
/Z-\alpha (C^z+C/2)-(1-\alpha)/(2Z)] \nonumber\\
&&-\alpha\lambda J_{\perp {\rm eff}}
[\epsilon (C_{\perp }^{z}+\chi_{\perp}^{z})+\epsilon_{\perp}
(C_{\perp}+\epsilon\chi_{\perp})/2], \nonumber\\
A_{3}=&&\lambda^{2}[\alpha (C^{z}+\epsilon^{2}C/2)+(1-\alpha)
(1+\epsilon^{2})/(4Z)-\alpha\epsilon (\chi/2+\epsilon\chi^{z})/Z]
\nonumber\\
&&+\alpha\lambda J_{\perp {\rm eff}}[\epsilon\epsilon_{\perp}
C_{\perp}+2C_{\perp}^{z}]+J_{\perp {\rm eff}}^{2}
(\epsilon_{\perp}^{2}+1)/4, \nonumber\\
X_{1}=&&\alpha\lambda J_{\perp {\rm eff}}[(\epsilon_{\perp}\chi +
\epsilon\chi_{\perp})/2+\epsilon\epsilon_{\perp}(\chi_{\perp}^{z}+
\chi^{z})], \nonumber\\
X_{2}=&&-\alpha\lambda J_{\perp {\rm eff}}[\epsilon
\epsilon_{\perp}\chi/2+\epsilon_{\perp}(\chi^{z}+C_{\perp}^{z})+
\epsilon C_{\perp}/2]-\epsilon_{\perp}J_{\perp {\rm eff}}^{2}/2,
\end{eqnarray}
the spinon correlation functions $\chi =\langle S_{ai}^{+}
S_{ai+\hat{\eta}}^{-}\rangle$, $\chi^{z}=\langle S_{ai}^{z}
S_{ai+\hat{\eta}}^{z}\rangle$, $\chi_{\perp}=\langle S_{1i}^{+}
S_{2i}^{-}\rangle$, $\chi_{\perp}^{z}=\langle S_{1i}^{z}S_{2i}^{z}
\rangle$, $C=(1/Z^2)\sum_{\hat{\eta}\hat{\eta^{\prime}}}\langle
S_{ai+\hat{\eta}}^{+}S_{ai+\hat{\eta^{\prime}}}^{-}\rangle$, and
$C^{z}=(1/Z^2)\sum_{\hat{\eta}\hat{\eta ^{\prime }}}\langle
S_{ai+\hat{\eta}}^{z}S_{ai+\hat{\eta^{\prime }}}^{z}\rangle$,
$C_{\perp}=(1/Z)\sum_{\hat{\eta}}\langle S_{2i}^{+}
S_{1i+\hat{\eta}}^{-}\rangle$, and $C_{\perp}^{z}=(1/Z)
\sum_{\hat{\eta}}\langle S_{1i}^{z}S_{2i+\hat{\eta}}^{z}\rangle$.
In order to satisfy the sum rule for the correlation function
$\langle S_{ai}^{+}S_{ai}^{-}\rangle=1/2$ in the absence of AF
long range order (AFLRO), a decoupling parameter $\alpha $ has
been introduced in the MF calculation, which can be regarded as
the vertex correction. \cite{sp97} All these parameters have been
determined self-consistently, as done in the single layer case.
\cite{yuan}

At half-filling, the $t$-$J$ model is reduced to the  Heisenberg
AF model, and the AFLRO gives rise to a commensurate peak at
$[1/2,1/2]$ (hereafter we use the units of $[2\pi ,2\pi ]$). In
Fig. 1, we plot DSSF $S({\bf k},\omega)$ in the ($k_{x},k_{y}$)
plane at doping $p=0.06$, temperature $T=0.1J$ and energy $\omega=
0.35J$ for $t/J=2.5$, $t_{\perp }/t=0.25$, and $J_{\perp}/J=0.25$,
which shows that a commensurate-IC transition in the spin
fluctuation pattern occurs with doping. At low energies and lower
temperatures, the commensurate peak close to  half-filling is
split into four peaks at $[(1\pm \delta )/2,1/2]$ and
$[1/2,(1\pm \delta )/2]$. The calculated DSSF $S({\bf k},\omega)$
has been used to extract the doping dependence of the IC parameter
$\delta (p)$, defined as the deviation of the peak position from
the AF wave vector $[1/2, 1/2]$, and the result is shown in Fig. 2
in comparison with the experimental data\cite{dai01} taken on YBCO
(inset). $\delta (p)$ increases initially with the hole
concentration in the low doping regime, but it saturates quickly
at higher dopings, in semi-quantitative agreement with the
experimental data. \cite{dai01} Apparently, there is a substantial
difference between theory and experiment, namely the saturation
occurs at $p=0.10$ in experiment, while the calculation
anticipates it already at $p\approx 0.05$. However, upon a closer
examination one sees immediately that the main difference is due
to the appearance of ICAF at too low dopings in the theoretical
consideration. The actual range of rapid growth of IC parameter
$\delta (p)$ with doping $p$ (around $4\sim 5\%$) is very similar
in theory and experiment.

For considering the resonance at relatively high energy we have
made a series of scans for $S({\bf k},\omega )$ at different
energies, and the result for doping $p=0.06$, $t/J=2.5$,
$t_{\perp }/t=0.25$, $J_{\perp}/J=0.25$ at $T=0.1J$ and
$\omega =0.5J$ is shown in Fig. 3. Comparing it with Fig. 1 for
the same set of parameters except for $\omega =0.35J$, we see that
IC peaks are energy dependent, {\it i.e.}, although these magnetic
scattering peaks retain the IC pattern at $[(1\pm \delta )/2,1/2]$
and $[1/2,(1\pm \delta )/2]$ in low energies, the positions of IC
peaks move towards  $[1/2,1/2]$ with increasing energy, and then
the $[1/2,1/2]$ resonance peak appears at relatively high energy
($\omega_{r}=0.5J$). To show this point clearly, we plot the
evolution of the magnetic scattering peaks with energy in Fig. 4.
For comparison, the experimental result \cite{bour} of
YBa$_2$Cu$_3$O$_{6+x}$ with $x=0.85$ $(p\approx 0.14)$ for the SC
state is shown in the same figure. A similar experimental result
\cite{ybic} has also been obtained for YBa$_2$Cu$_3$O$_{6+x}$ with
$x=0.7$ $(p\approx 0.12)$. Although these experimental data were
obtained below $T_{c}$, they also hold  for the normal state in
the underdoped regime $x\leq 0.7$ $(p\leq 0.12)$.\cite{dai01}
The anticipated position $\omega_{r}=0.5J\approx 50$ mev
\cite{shamoto} is not too far from the peak $\approx 30$mev
$\sim 37$mev observed in underdoped YBCO. \cite{dai01} Moreover,
the resonance energy $\omega_{r}$ is proportional to $p$ at small
dopings. We have also made a series of scans for
$S({\bf k},\omega)$ at different temperatures, and found both IC
peaks and resonance peak are broadened and suppressed with
increasing temperature, and tend to vanish at high temperatures.
This reflects that the spin excitations are rather sharp in
momentum space at low temperatures, compared with  the linewidth,
and the inverse lifetime increases with increasing temperature.
Our result is in qualitative agreement with experiments.
\cite{dai01}

Now we turn to discuss the integrated spin response, which is
manifested by the integrated dynamical spin susceptibility, and
can be expressed as,
\begin{eqnarray}
I(\omega,T)=(1/N)\sum_{k}\chi^{\prime\prime}({\bf k},\omega),
\end{eqnarray}
where the dynamical spin susceptibility is related to DSSF by the
fluctuation-dissipation theorem as, $\chi^{\prime\prime}({\bf k},
\omega)=(1-e^{-\beta\omega})S({\bf k},\omega)$. The results of
$I(\omega,T)$ at doping $p=0.06$ in $t/J=2.5$, $t_{\perp}/t=0.25$,
and $J_{\perp}/J=0.25$ with $T=0.1J$ (solid line) and $T=0.2J$
(dashed line) are plotted in Fig. 5 in comparison with the
experimental data \cite{bir} taken from YBa$_{2}$Cu$_{3}$O$_{6+x}$
(inset), where the dotted line is the function $\sim{\rm arctan}
[a_{1}\omega/T+a_{3}(\omega/T)^{3}]$ with $a_{1}=6.6$, and
$a_{3}=3.9$. These results show that $I(\omega,T)$ is almost
constant for $\omega/T >1$ and then begin to decrease with
decreasing $\omega/T$ for $\omega/T <1$. It is quite remarkable
that the integrated susceptibility in the bilayer cuprates shows
the same universal behavior as in the case of the single layer
cuprates, \cite{yuan} and is scaled approximately as
$I(\omega,T)\propto{\rm arctan}[a_{1}\omega/T+a_{3}(\omega/T)^{3}
]$. This result is consistent with experiments. \cite{bir}

The DSSF in Eq. (3) has a well-defined resonance character,
where $S({\bf k},\omega)$ exhibits peaks when the incoming
neutron energy $\omega$ is equal to the renormalized spin
excitation $E_{k}^2=(\omega_{k}^{(1)})^{2}+B_{k}^{(1)}{\rm Re}
\Sigma_{LT}^{(s)}({\bf k},E_{k})$, {\it i.e.},
$W({\bf k}_{c},\omega)\equiv [\omega^{2}-
(\omega_{k_{c}}^{(1)})^{2}-B_{k_{c}}^{(1)}{\rm Re}
\Sigma_{LT}^{(s)}({\bf k}_{c},\omega)]^{2}=(\omega^{2}-
E_{k_{c}}^{2})^{2}\sim 0$ for certain critical wave vectors
${\bf k}_{c}$. The height of these peaks is determined by the
imaginary part of the spinon self-energy $1/{\rm Im}
\Sigma_{LT}^{(s)}({\bf k}_{c},\omega)$. This renormalized spin
excitation is doping and energy dependent. Since ${\rm Re}
\Sigma_{LT}^{(s)}({\bf k},\omega)={\rm Re}\Sigma_L^{(s)}({\bf k},
\omega )+{\rm Re}\Sigma _{T}^{(s)}({\bf k},\omega)$ with ${\rm Re}
\Sigma_{L}^{(s)}({\bf k},\omega)<0$ and ${\rm Re}\Sigma_{T}^{(s)}
({\bf k},\omega)>0$,  there is a competition between ${\rm Re}
\Sigma_{L}^{(s)}({\bf k},\omega)$ and ${\rm Re}\Sigma_{T}^{(s)}
({\bf k},\omega)$, which comes entirely from the bilayer band
splitting.\cite{feng} At low energies the main contribution to
${\rm Re}\Sigma_{LT}^{(s)}({\bf k},\omega)$ comes from ${\rm Re}
\Sigma_{L}^{(s)}({\bf k},\omega)$, then ICAF emerges, where the
essential physics is almost the same as in single layer cuprates,
and  detailed explanations have been given in  Ref. \cite{yuan}.
Near half-filling, the spin excitations are centered around the AF
wave vector [1/2, 1/2], so the commensurate AF peak appears there.
Upon doping, the holes disturb the AF background. Within the
fermion-spin framework, as a result of self-consistent motion of
holons and spinons, ICAF is developed beyond certain critical
doping, which means, the low-energy spin excitations drift away
from the AF wave vector, or the zero of
$W({\bf k}_{\delta},\omega)$ is shifted from $[1/2,1/2]$ to
${\bf k}_{\delta}$, where the physics is dominated by the spinon
self-energy ${\rm Re}\Sigma_{L}^{(s)}({\bf k},\omega)$
renormalization due to holons. In this sense, the mobile holes are
the key factor leading to ICAF. However, ${\rm Re}\Sigma_{T}^{(s)}
({\bf k},\omega)$ cancels out most contributions from ${\rm Re}
\Sigma_{L}^{(s)}({\bf k},\omega)$ at relatively high energy, then
the anomalous $[1/2,1/2]$ resonance reappears. Therefore the
bilayer band splitting plays a crucial role in  giving rise to the
resonance. What we calculate is the acoustic spin excitation
with modulations in the $c$-direction $\propto \sin^{2}
(\pi z_{Cu} L)$, where $z_{Cu}$ is the distance between two
nearest Cu layers, $L$ the $c$-axis coordinate in the  reciprocal
space. This reflects the {\it antiferromagnetic coupling} between
layers, and it is fully confirmed by experiments.
\cite{ybic,dai01,bour}  

In conclusion  we have  shown that if the strong spinon-holon
interaction and bilayer interactions are taken into account, the
$t$-$J$ model per se can correctly reproduce all main features of
INS experiments in the {\it normal state} in underdoped bilayer
cuprates, including the doping  and energy dependence of ICAF at
low energies and $[1/2,1/2]$ resonance at relatively high energy.
In fact the ICAF peaks converge to the commensurate resonance, as
the energy is increased. In our opinion, the difference of AF
fluctuation behavior between LSCO and YBCO (BSCO) is not due to
the presence/absence of stripes, but rather  because of the
single/double layer structure. Of course, this has to be checked
by further experiments. It is possible that at some particular
energy, a strong commensurate resonance peak coexists with the
weaker IC features as shown in Fig. 3.

After submitting this paper, we became aware of the recent INS
measurements\cite{he} providing evidence for a sharp commensurate
resonance peak below $T_{c}$ in the single layer cuprate
Tl$_{2}$Ba$_{2}$Cu$_{6+\delta}$ near optimal doping. However,
above $T_{c}$, the experimental scans show a featureless
background that gradually decreases in an energy- and
momentum-independent fashion as the temperature is lowered. The
INS in the SC state has not been considered so far within the
fermion-spin approach, and we need to extend our studies for both
single layer\cite{yuan} and bilayer cases to the SC state, where
the holon Cooper pairs are formed, and the spinon self-energy is
originating from both normal and {\it anomalous} holon bubbles.
Hence the renormalized spin excitation in the SC state is very
much different from that in the {\it normal state}, and it may
be related to the magnetic peaks detected in the SC state. These
and other related issues are under investigation now. On the other
hand, we emphasize that although the simple $t$-$J$ model can not
be regarded as a comprehensive  model for the quantitative
comparison with the doped cuprates, our present results for the
{\it normal state} are in semi-quantitative agreement with the
major experimental observations in the {\it normal state} of the
underdoped bilayer cuprates. \cite{dai01,bour,bir}.

\acknowledgments
This work was supported by the National Natural Science Foundation
of China under Grant Nos. 10074007, 10125415, and 90103024.

\begin{figure}[tbp]
\caption{The dynamical spin structure factor in the $(k_x,k_y)$
plane at doping $p=0.06$, temperature $T=0.1J$ and energy
$\omega =0.35J$ for  $t/J=2.5$, $t_{\perp }/t=0.25$, and
$J_{\perp }/J=0.25$. A quasiparticle damping $\Gamma=0.01J$ has
been used in all results presented.}
\end{figure}

\begin{figure}[tbp]
\caption{The doping dependence of the incommensurability
$\delta(p)$ of the antiferromagnetic fluctuations. Inset: the
experimental results on YBCO taken from
Ref. \protect\cite{dai01}.}
\end{figure}

\begin{figure}[tbp]
\caption{The dynamical spin structure factor in the
$(k_{x},k_{y})$ plane at $p=0.06$ for $t/J=2.5$,
$t_{\perp}/t=0.25$, $J_{\perp}/J=0.25$ and $\omega=0.5J$ at
$T=0.1J$. }
\end{figure}

\begin{figure}[tbp]
\caption{The energy dependence of the  position of the
incommensurate peaks at $p=0.06$ and $T=0.1J$ for $t/J=2.5$,
$t_{\perp}/t=0.25$, and $J_{\perp}/J=0.25$ (left ordinates) vs
the experimental results on YBa$_{2}$Cu$_{3}$O$_{6.85}$ in the
superconducting state taken from Ref. \protect\cite{bour} (right
ordinates). }
\end{figure}

\begin{figure}[tbp]
\caption{The integrated susceptibility at $p=0.06$ for $t/J=2.5$,
$t_{\perp}/t=0.25$, and $J_{\perp}/J=0.25$ in $T=0.1J$ (solid
line) and $T=0.2J$ (dashed line). The dotted line is the function
$b_{1}{\rm arctan}[a_{1}\omega/T+a_{3}(\omega/T)^{3}]$ with
$a_{1}=6.6$ and $a_{3}=3.9$. Inset: the experimental result on
YBa$_{2}$Cu$_{3}$O$_{7-x}$ taken from Ref. \protect\cite{bir}.}
\end{figure}


\begin{references}
\bibitem{rev} For reviews, see M.A.Kastner, R.J. Birgeneau, G.
Shiran, Y. Endoh, Rev. Mod. Phys. {\bf 70}, 897 (1998); A.P.
Kampf, Phys. Rep. {\bf 249}, 219 (1994).

\bibitem{yamada} K. Yamada, C.H. Lee, K. Kurahashi, J. Wada, S.
Wakimoto, S. Ueki, H. Kimura, Y. Endoh, S. Hosoya, and G. Shirane,
Phys. Rev. B {\bf 57}, 6165 (1998), and references therein.

\bibitem{aeppli} G. Aeppli, T.E. Mason, S.M. Hayden, H.A. Mook,
and J. Kulda, Science {\bf 278}, 1432 (1997).

\bibitem{ros91} J. Rossat-Mignod, L.P. Regnault, P. Bourges, P.
Burlet, J. Bossy, J.Y. Henri, and G. Lapertot, Physica C
{\bf 185-189}, 86 (1991).

\bibitem{res} H.A. Mook, M. Yethiraj, G. Aeppli, T.E. Mason, and
T. Armstrong, Phys. Rev. Lett. {\bf 70}, 3490 (1993); H.F. Fong,
B. Keimer, P.W. Anderson, D. Reznik, F. Do\~gan, and I.A. Aksay,
Phys. Rev. Lett. {\bf 75}, 316 (1995); P. Bourges, L.P. Regnault,
Y. Sidis, and C. Vettier, Phys. Rev. B {\bf 53}, 876 (1996).

\bibitem{resunder} P. Dai, M. Yethiraj, H.A. Mook, T.B. Lindemer,
and F. Do\~gan, Phys. Rev. Lett. {\bf 77}, 5425 (1996); H.F. Fong,
B. Keimer, F. Do\~gan, and I.A. Aksay, Phys. Rev. Lett. {\bf 78},
713 (1997); P. Bourges, L.P. Regnault, Y. Sidis, J. Bossy, P.
Burlet, C. Vettier, J.Y. Henry, and M. Couach, Europhys. Lett.
{\bf 38}, 313 (1997).

\bibitem{bsco} H.F. Fong, P. Bourges, Y. Sidis, L.P. Regnault, A.
Ivanov, G.D. Gu,  N. Koshizuka, and B. Keimer, Nature {\bf 398},
588 (1999); H. He, Y. Sidis, P. Bourges, G.D. Gu, A. Ivanov, N.
Koshizuka, B. Liang, C.T. Lin, L.P. Regnault, E. Schoenherr,
and B. Keimer, Phys. Rev. Lett. {\bf 86}, 1610 (2001).

\bibitem{ybic} H.A. Mook, P. Dai, S.M. Hayden, G. Aeppli, T.J.
Perring, and F. Do\~gan, Nature {\bf 395}, 580 (1998); M. Arai,
T. Nishijima, Y. Endoh, T. Egami, S. Tajima, K. Tomimoto, Y.
Shiohara, M. Takahashi, A. Garret, and S.M. Bennington, Phys.
Rev. Lett. {\bf 83}, 608 (1999); P. Dai, H.A. Mook, S.M. Hayden,
G. Aeppli, T.J. Perring, R.D. Hunt, and F. Do\~gan, Science
{\bf 284}, 1344 (1999).

\bibitem{dai01} P. Dai, H.A. Mook, R.D. Hunt, and F. Do\~gan,
Phys. Rev. B{\bf 63}, 54525 (2001), and references therein.

\bibitem{bour} P. Bourges, Y. Sidis, H.F. Fong, L.P. Regnault,
J. Bossy, A. Ivanov, and B. Keimer, Science {\bf 288}, 1234
(2000); P. Bourges, B. Keimer, L.P. Regnault, and Y. Sidis,
cond-mat/0006085.

\bibitem{nesting} See, {\it e.g.}, N. Bulut, D. Hone, D.J.
Scalapino, and N.E. Bickers, Phys. Rev. Lett {\bf 64}, 2723
(1990); Q. Si, Y. Zha, K. Levin, and J.P. Lu, Phys. Rev. B
{\bf 47}, 9055 (1993); T. Tanamoto, H. Kohno, and H. Fukuyama,
J. Phys. Soc. Jpn. {\bf 63}, 2739 (1994).

\bibitem{earthe} See, {\it e.g.}, G. Blumberg, B.P. Stojkovic,
and M.V. Klein, Phys. Rev. B {\bf 52}, 15741 (1995); D.Z. Liu,
Y. Zha, and K. Levin, Phys. Rev. Lett. {\bf 75}, 4130 (1995)
and more references in Ref. \protect\cite{brink}.

\bibitem{stripe} See, {\it e.g.}, J. Zaanen and O. Gunnarsson,
Phys. Rev. B {\bf 40}, 7391 (1989); D. Poilblanc and T.M. Rice,
Phys. Rev. B {\bf 39}, 9749 (1989).

\bibitem{latthe} D.K. Morr and D. Pines, Phys. Rev. Lett.
{\bf 81}, 1086 (1998); F. Onufrieva and P. Pfeuty,
cond-mat/9903097; A. Abanov and A.V. Chubukov, Phys. Rev. Lett.
{\bf 83}, 1652 (1999). J.Y. Kao, Q. Si, and K. Levin,, Phys. Rev.
B {\bf 61}, 11898 (2000); M.R. Norman, Phys. Rev. B {\bf 61},
14751 (2000).

\bibitem{brink} J. Brinckmann and P.A. Lee, Phys. Rev. Lett.
{\bf 82}, 2915 (1999); {\it ibid}, cond-mat/0107138.

\bibitem{so5} E. Demler and S.C. Zhang, Phys. Rev. Lett. {\bf 75},
4126 (1995); E. Demler, H. Kohno, and S.C. Zhang, Phys. Rev. B
{\bf 58}, 5719 (1998).

\bibitem{yin} L. Yin, S. Chakravarty, and P.W. Anderson, Phys.
Rev. Lett. {\bf 78}, 3559 (1997).

\bibitem{feng} D.L. Feng, N.P. Armitage, D.H. Lu, A. Damascelli,
J.P. Hu, P. Bogdanov, A. Lanzara, F. Ronning, K.M. Shen, H.
Eisaki, C. Kim, Z.X. Shen, J.-i. Shimoyama, and K. Kishio, Phys.
Rev. Lett. {\bf 86}, 5550 (2001).

\bibitem{sp94} Shiping Feng, Z.B. Su, and L. Yu, Phys. Rev. B
{\bf 49}, 2368 (1994); Mod. Phys. Lett. B{\bf 7}, 1013 (1993).

\bibitem{yuan} Feng Yuan, Shiping Feng, Zhao-Bin Su, and Lu Yu,
Phys. Rev. B {\bf 64}, 224505 (2001); Shiping Feng and Zhongbing
Huang, Phys. Rev. B {\bf 57}, 10328 (1998).

\bibitem{sp97} Shiping Feng and Yun Song, Phys. Rev. B {\bf 55},
642 (1997); J. Kondo and K. Yamaji, Prog. Theor. Phys. {\bf 47},
807 (1972).

\bibitem{shamoto} S. Shamoto, M. Sato, J.M. Tranquada, B.J.
Sternlib, and G. Shirane, Phys. Rev. B {\bf 48}, 13817 (1993).

\bibitem{bir} R.J. Birgeneau, R.W. Erwin, P.G. Gehring, B.
Keimer, M.A. Kastner, M. Sato, S. Shamoto, G. Shirane, and J.M.
Tranquada, Z. Phys. B{\bf 87}, 15 (1992); B.J. Sternlieb, G.
Shirane, J.M. Tranquada, M. Sato, and S. Shamoto, Phys. Rev. B
{\bf 47}, 5320 (1993).

\bibitem{he} H. He, P. Bourges, Y. Sidis, C. Ulrich, L.P.
Regnault, S. Pailh\'es, N.S. Berzigiarova, N.N. Kolesnikov, and
B. Keimer, Science {\bf 295}, 1045 (2002).

\end{references}
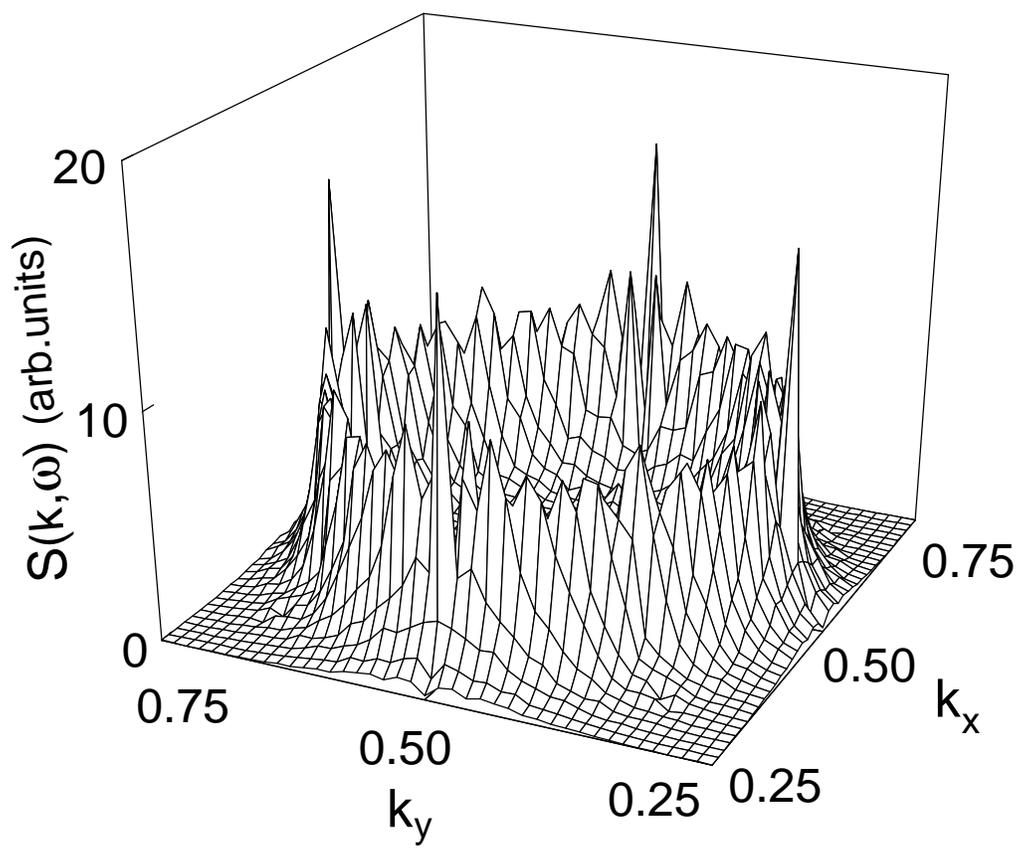
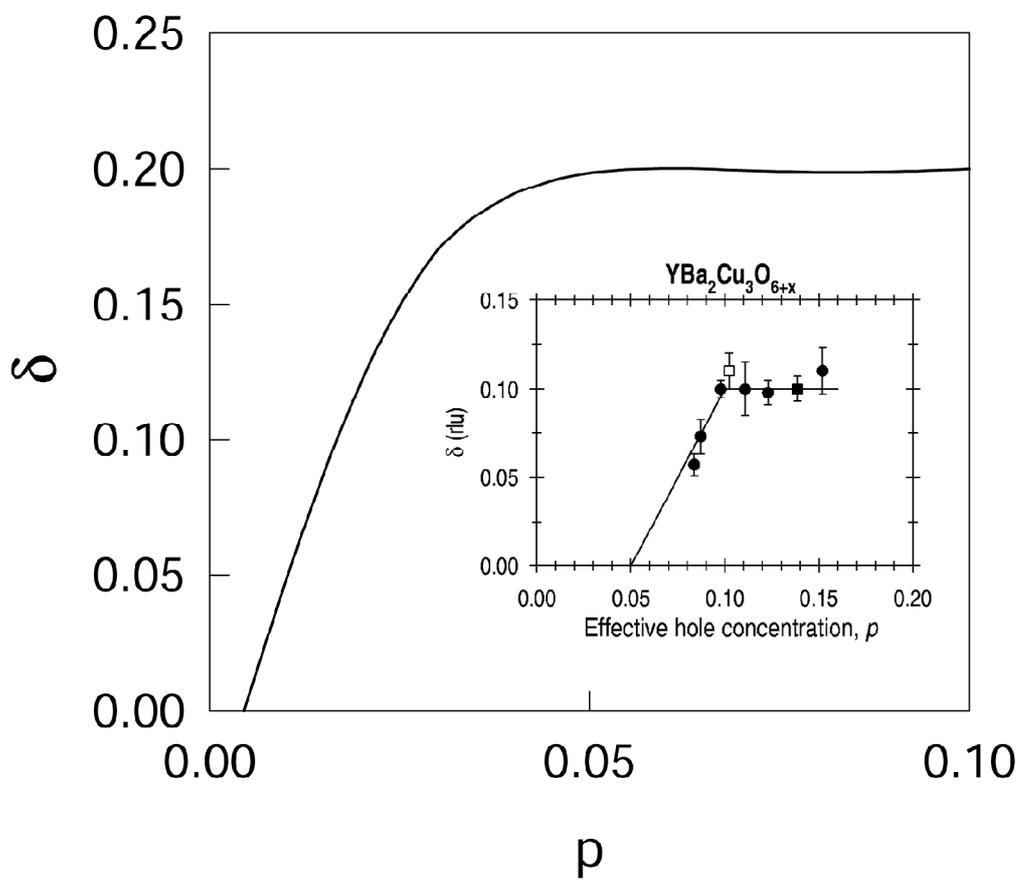
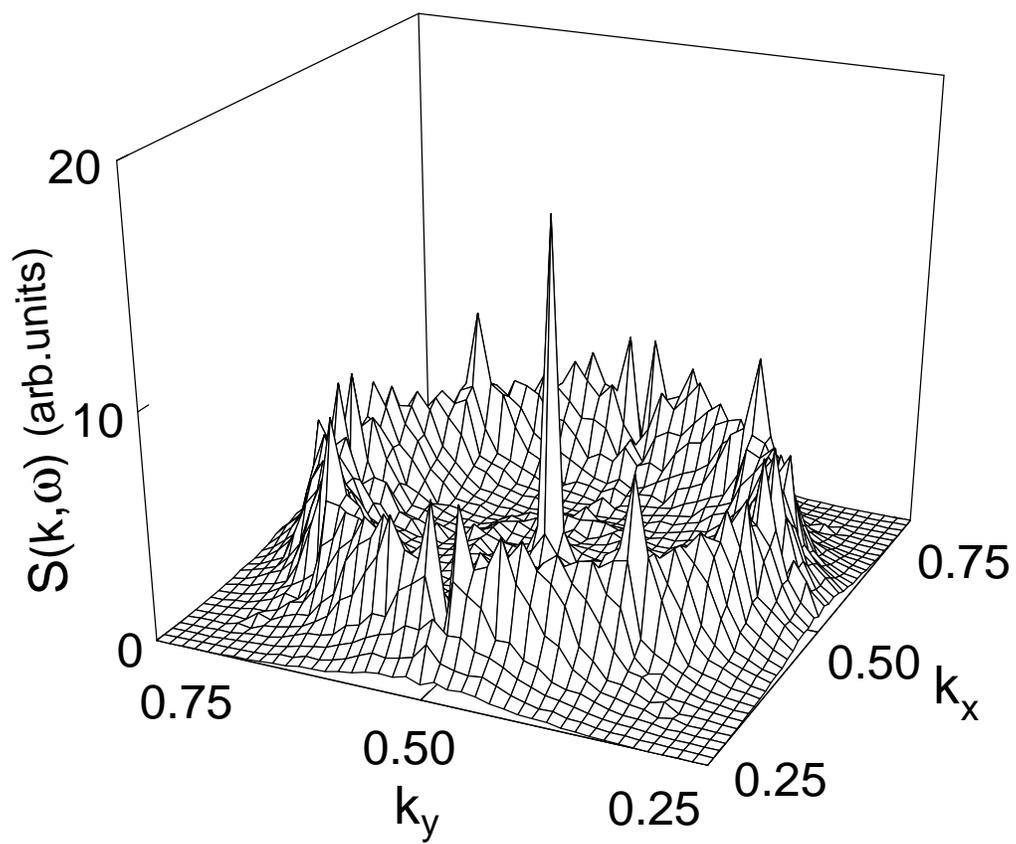
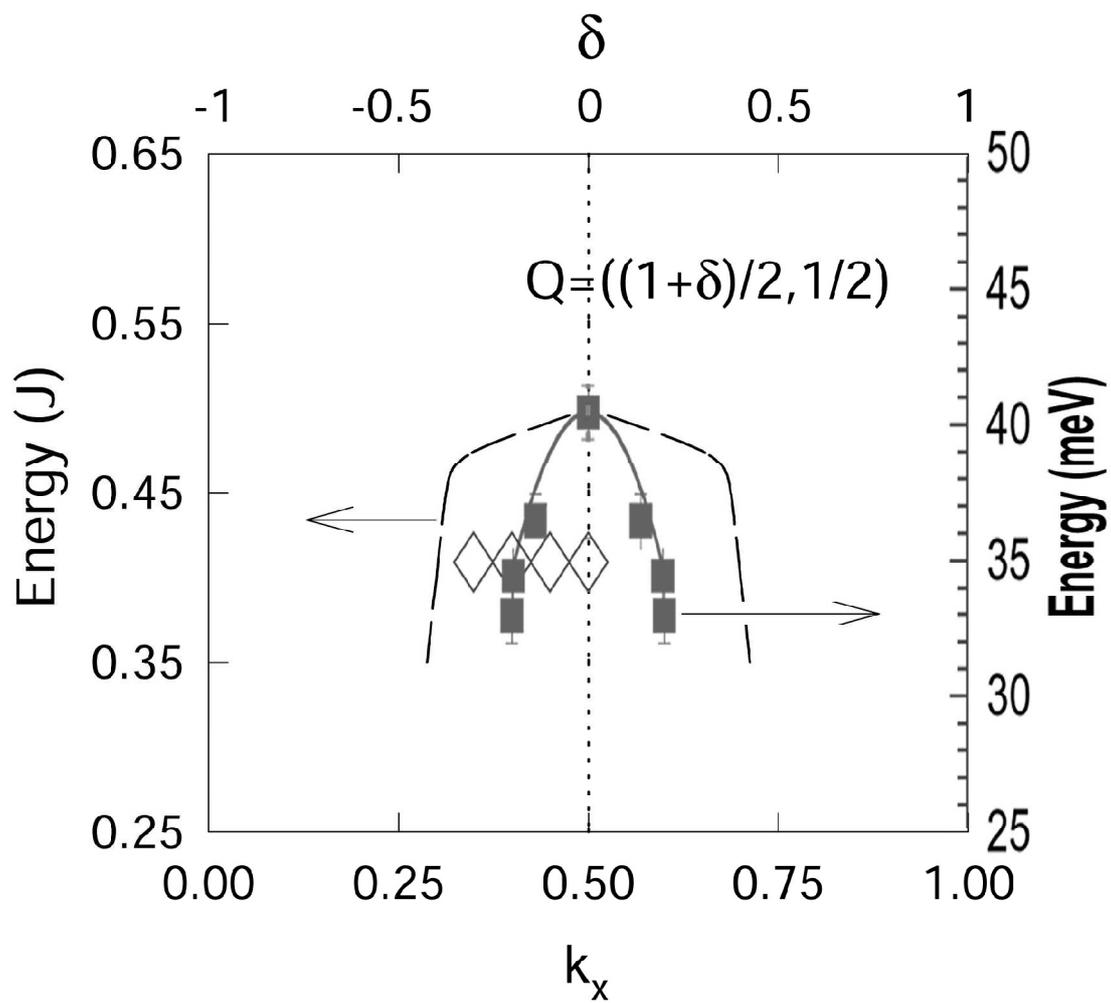
\end{document}